\documentclass[graybox]{svmult}

\usepackage{amsfonts}
\usepackage{amssymb}
\usepackage{mathptmx}       % selects Times Roman as basic font
\usepackage{helvet}         % selects Helvetica as sans-serif font
\usepackage{courier}        % selects Courier as typewriter font
\usepackage{type1cm}        % activate if the above 3 fonts are
\usepackage{makeidx}         % allows index generation
\usepackage{graphicx}        % standard LaTeX graphics tool
\usepackage{multicol}        % used for the two-column index
\usepackage[bottom]{footmisc}% places footnotes at page bottom

\makeindex             % used for the subject index
\begin{document}

\title*{Increased signaling entropy in cancer requires the scale-free property of protein interaction networks}
\author{Andrew E. Teschendorff, Chris R. S. Banerji, Simone Severini, Reimer K\"uhn and Peter Sollich}
\institute{Andrew E Teschendorff \at CAS Key Lab of Computational Biology, CAS-MPG Partner Institute for Computational Biology, Chinese Academy of Sciences, Shanghai Institute for Biological Sciences, 320 Yue Yang Road, Shanghai 200031, China. \at Statistical Cancer Genomics, Paul O'Gorman Building, UCL Cancer Institute, University College London, London WC1E 6BT, UK. \email{a.teschendorff@ucl.ac.uk} \and Chris Banerji \at Statistical Cancer Genomics, Paul O'Gorman Building, UCL Cancer Institute, University College London, London WC1E 6BT, UK. \and Simone Severini \at Dept.of Computer Science, University College London, London WC1E 6BT, UK. \and Reimer K\"uhn \at Department of Mathematics, King's College London, London WC2R 2LS, UK. \and Peter Sollich \at Department of Mathematics, King's College London, London WC2R 2LS, UK.}

\titlerunning{Entropy, Cancer and Scale-Freeness}
\authorrunning{Teschendorff AE et al}
\maketitle

\abstract{One of the key characteristics of cancer cells is an increased phenotypic plasticity, driven by underlying genetic and epigenetic perturbations. However, at a systems-level it is unclear how these perturbations give rise to the observed increased plasticity. Elucidating such systems-level principles is key for an improved understanding of cancer. Recently, it has been shown that signaling entropy, an overall measure of signaling pathway promiscuity, and computable from integrating a sample's gene expression profile with a protein interaction network, correlates with phenotypic plasticity and is increased in cancer compared to normal tissue. Here we develop a computational framework for studying the effects of network perturbations on signaling entropy. We demonstrate that the increased signaling entropy of cancer is driven by two factors: (i) the scale-free (or near scale-free) topology of the interaction network, and (ii) a subtle positive correlation between differential gene expression and node connectivity. Indeed, we show that if protein interaction networks were random graphs, described by Poisson degree distributions, that cancer would generally not exhibit an increased signaling entropy. In summary, this work exposes a deep connection between cancer, signaling entropy and interaction network topology.}

\vspace{0.5cm}


\section*{Introduction}
One of the key features of cancer is an increased cellular plasticity, mediated by an increased promiscuity in signaling patterns, and driven by underlying genetic and epigenetic aberrations which cause a fundamental rewiring of the intracellular signaling network \cite{Califano2011,Dutkowski2011,Creixell2012,Schramm2010,Ideker2010,Ideker2012,Csermely2013,Pisco2013}. Every aberration found in a cancer cell can be thought of as a perturbation if the aberration affects the gene functionally. Such perturbations can be classed as activating, if they result in an increased functional activity of the gene (e.g. amplification and overexpression of {\it ERBB2} in breast cancer), or inactivating, if it compromises gene function (e.g. silencing through promoter DNA methylation). Whilst the effect of certain specific perturbations on gene function can be predicted, it is much less clear how individual perturbations affect the cellular phenotype as a whole, since this depends on the collective nature of the other aberrations that are present in the same cell. Predicting the net effect of multiple perturbations in a signaling network is hard due to complex effects such as pathway redundancy and epistasis \cite{Creixell2012,Ideker2012}. Moreover, in the context of cancer, although the effect of specific aberrations on cell function is known, it is yet unclear how individual cancer perturbations may contribute to the observed increased signaling promiscuity and phenotypic plasticity.\\
One way to approach this challenge computationally, is to anchor the analysis on global measures which capture salient features of the cellular phenotype, and which are computable from, say, a sample's molecular profile (e.g. a sample's gene expression profile). Here we are particularly interested in measuring signaling promiscuity since evidence is mounting that this underlies a sample's phenotypic plasticity \cite{Pisco2013}. In previous work we have started to explore a measure which approximates intra-sample signaling promiscuity, and which is known as network signaling entropy \cite{West2012,Banerji2013,Teschendorff2014}. Signalling entropy is computed from integrating a sample's genome-wide gene expression profile with a protein interaction network and, as shown by us, provides a surprisingly good estimate of a sample's height in Waddingtons's differentiation landscape, with human embryonic stem cells (hESCs) exhibiting the highest levels of entropy \cite{Banerji2013}. Indeed, signalling entropy was able to discriminate cellular samples according to their differentiation potential within distinct lineages, including hematopoietic, mesenchymal and neural lineages, and with terminally differentiated cells within these lineages exhibiting the lowest levels of entropy \cite{Banerji2013}. Importantly, signaling entropy was also found to be higher in cancer compared to normal tissue, consistent with the view that cancer cells represent a more undifferentiated stem-cell like state, characterised by an increase in phenotypic plasticity \cite{Banerji2013,Pisco2013}. \\ 
Given that increased signaling entropy is such a robust and characteristic feature of differentiation potency and cancer, and that it is also amenable to computation \cite{West2012,Banerji2013}, it is of great theoretical and biological interest to study the changes in entropy caused by cellular network perturbations. In the context of cancer, two well-known network perturbations are the overexpression and underexpression of oncogenes and tumour suppressor genes, respectively, and although these perturbations are known to result in the uncontrolled activation of cell-growth and cell-proliferation pathways, it remains unclear how these perturbations affect signalling promiscuity. In order to deepen our understanding, we here decided to study the effect of such perturbations on signaling entropy, using both simulated and real data, and using a variety of different network types in order to assess the impact of network topology. Specifically, we consider Erdos-Renyi random (Poisson) graphs \cite{Erdos1959}, scale free networks \cite{Barabasi1999}, as well as real protein-protein interaction (PPI) networks \cite{Prasad2009,Cerami2011,Rolland2014}. In doing so, we discover that in Poisson networks, perturbations (be they activating or inactivating) lead to reductions in the global entropy, but that this is not true for scale-free and more realistic PPI networks. In networks exhibiting a scale-free, or near scale-free topology, we show that gene expression perturbations affecting hubs exhibit a striking bi-modality, leading to increases or decreases in the global entropy rate depending on the directionality of the expression change. We further expose a subtle yet significantly positive correlation between differential gene expression in cancer and node-degree, which we show drives the increased signaling promiscuity of cancer, but only if the underlying protein interaction network has a scale-free (or near scale-free) topology. Thus, this work makes a deep connection between a defining feature of the cancer phenotype, i.e. high signaling entropy, its differential gene expression pattern and the (near) scale-free topology of real PPI networks.\\
Although there are many studies on network perturbations, it is worth clarifying that the network perturbations and outcome of interest (i.e. the entropy rate) considered in this work are very different from the perturbations and outcomes of interest considered in previous studies \cite{Albert2000,Jeong2001,Serra-Musach2012,Kim2014,Wang2014}. Specifically, we consider network perturbations which only alter the local edge weights without altering the underlying network topology \cite{Teschendorff2010bmc,West2012,Banerji2013}. Moreover, our network perturbations can be both activating as well as inactivating, representing the two different types of cancer alterations affecting oncogenes and tumour suppressors, respectively. In contrast, much of the previous literature has dealt with the effects of removing specific nodes in unweighted networks \cite{Albert2000,Jeong2001}, a type of inactivating perturbation which alters the underlying network topology, focusing on tolerance and robustness as outcome measures \cite{Albert2000,Jeong2001,Serra-Musach2012,Wang2014}. Thus, from a network theoretical perspective, the important novel insights reported in this work are made possible by considering a novel type of network perturbation in the context of weighted networks defined by a stochastic matrix. We should also stress that our outcome of interest, signaling entropy, is a systems-level measure that is constructed from the genome-wide expression profile of a given sample, and therefore has little to do with the protein signaling disorder measures considered by other studies and which do not use gene expression data \cite{Haynes2006}.

\section*{Results}

\subsection*{Increased signaling entropy in cancer is driven by overexpression of hub genes}

In earlier work we demonstrated that signaling entropy, a measure of the signaling promiscuity in a cellular sample, is increased in cancer compared to normal tissue, irrespective of tissue type \cite{West2012,Banerji2013,Teschendorff2014}. This increased signaling entropy is consistent with the observed increased phenotypic plasticity of cancer cells (see e.g. \cite{Pisco2013}). Thus, increased signaling entropy has emerged as a cancer systems hallmark \cite{West2012,Teschendorff2014}. Signaling entropy is estimated as the entropy rate \cite{GomezGardenes2008} of a sample-specific stochastic matrix which models the signaling interactions in the sample ({\bf Appendix}). This stochastic matrix is computed by integrating the gene expression profile of the sample with a comprehensive PPI network, invoking the mass-action principle to define the edge-weights in the network ({\bf Appendix}). The mass-action principle is based on the assumption that two proteins, which have been reported to interact, are more likely to interact in a given sample if both are highly expressed in that sample.\\
Here we wanted to shed light on why, theoretically, we observe increased signaling entropy in cancer. We decided to use liver cancer as a model since liver represents a relatively homogeneous tissue, and is thus less affected by contaminating non-epithelial cells. We downloaded gene-normalised RNA-Seq data for a matched subset of 50 normal liver and 50 liver cancer samples from The Cancer Genome Atlas (TCGA). Confirming our earlier work using Affymetrix gene expression data \cite{West2012,Banerji2013,Teschendorff2014}, liver cancer exhibited a significantly higher signaling entropy rate compared to normal liver tissue ({\bf Fig.1A}).
\begin{figure}[ht]
\begin{center}
% For example, with the graphicx style use
\includegraphics[scale=0.75]{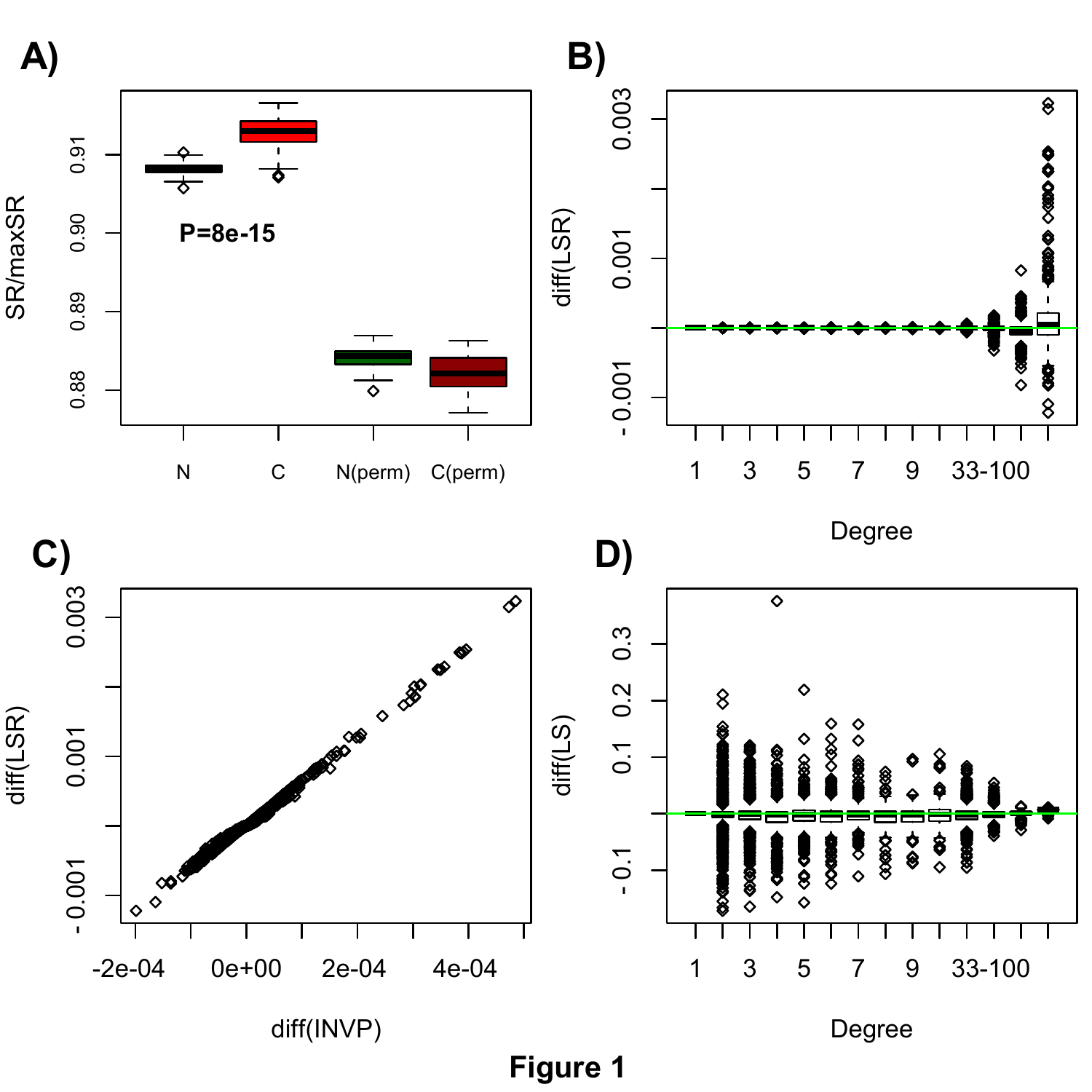}
%
% If no graphics program available, insert a blank space i.e. use
%\picplace{5cm}{2cm} % Give the correct figure height and width in cm
%
\caption{{\bf Increased entropy in liver cancer is driven by increased entropy at hubs:} {\bf A)} Boxplots comparing the entropy rate (SR) of 50 normal liver samples (N) to 50 matched liver cancer specimens (C), derived from RNA-Seq data of the TCGA consortium. P-value is from a one-tailed Wilcoxon rank sum test, testing the hypothesis that entropy rate is higher in cancer. Also shown is the SR between normal and liver cancer for a case where the gene expression profiles were randomly permuted (perm) over the interaction network. Observe how the difference in the SR between normal and cancer is reduced and even takes an opposite directionality, demonstrating that the interplay between gene expression changes and network topology is dictating the higher signaling entropy in cancer. {\bf B)} Boxplots showing the change in the mean local entropy rate (LSR) ($\langle\pi_iLS_i\rangle_C - \langle\pi_iLS_i\rangle_N$) between normal and cancer of each node (gene) as a function of node degree, positive values indicating higher values in cancer. {\bf C)} Scatterplot of the differential change in the mean local entropy rate against the differential change in the mean invariant measure (INVP) ($\langle\pi_i\rangle_C - \langle\pi_i\rangle_N$). Each data point is one node (gene). {\bf D)} Boxplots showing the change in the mean local entropy (LS) of each node (gene) ($\langle LS_i\rangle_C - \langle LS_i\rangle_N$) between normal and cancer, as a function of node degree.
}
\label{fig:1}       % Give a unique label
\end{center}
\end{figure}

Randomisation of the RNA-Seq profiles over the nodes in the network resulted in a significantly reduced difference in entropy rate between normal and cancer tissue ({\bf Fig.1A}),  indicating (as pointed out by us previously \cite{Teschendorff2014}) that the entropy increase in cancer is driven by a subtle interplay between specific gene expression changes and where these happen on the network. Specifically, we posited that the topological properties of the genes undergoing the largest changes in gene expression would be key features dictating the change in signalling entropy.\\
Since each gene $i$ contributes an amount $\pi_iLS_i$ to the entropy rate of a given sample ({\bf Appendix}), we computed for each gene the difference in the means of its local entropy rate, $\pi_iLS_i$, between normal and cancer tissue. In order to help interpretation, we also computed for each gene the difference in the means of the invariant measure $\pi_i$ between normal and cancer, as well as the difference in the average local entropy $LS_i$ ({\bf Appendix}). All these changes were assessed in relation to the connectivity of the genes in the network. We observed that the entropy rate increase in cancer is driven mainly by hubs, i.e. the nodes of highest degree in the network ({\bf Fig.1B}). Changes to the local entropy rates were driven by concomitant changes in the average invariant measure ({\bf Fig.1C}). Thus, hubs exhibited preferential increases in their average invariant measure, whilst also demonstrating positive increases in the average local entropy ({\bf Fig.1D}). Since the invariant measure value at a node $i$ represents the steady-state probability of finding a random walker at this node, the observed preferential increase in the invariant measure at hubs means that there is an increased signaling flux through these hub nodes in cancer.

\begin{figure}[ht]
\begin{center}
% For example, with the graphicx style use
\includegraphics[scale=0.8]{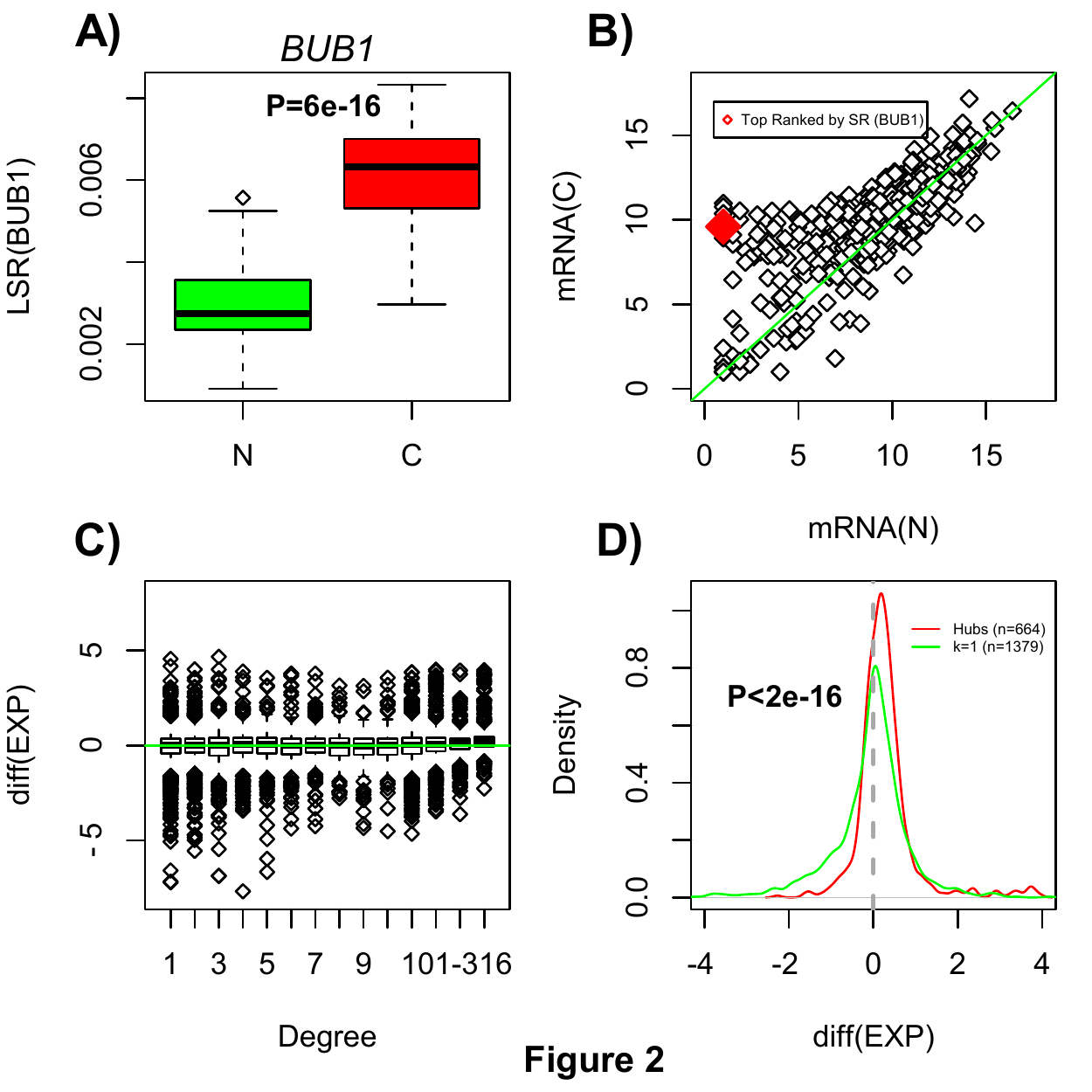}
%
% If no graphics program available, insert a blank space i.e. use
%\picplace{5cm}{2cm} % Give the correct figure height and width in cm
%
\caption{{\bf Preferential overexpression of hub genes in cancer:} {\bf A)} Boxplot showing the local entropy rate (LSR) against normal/cancer status, for the hub gene ({\it BUB1}) exhibiting the largest increase in the local entropy rate. P-value is from a Wilcoxon rank sum test. {\bf B)} Scatterplot of gene expression values between a representative normal (x-axis) and cancer (y-axis) sample for the gene showing the largest increase in the local entropy rate (gene {\it BUB1}, marked in red) and that of its neighbours in the PPI network (over 800 neighbours, shown in black). {\bf C)} Boxplot of the average difference in gene expression between normal and cancer (positive values indicate higher expression in cancer) against node-degree class. Observe how the highest-degree hubs show preferential increased expression in cancer, whereas the largest reductions in expression target low-degree nodes. {\bf D)} Density plot of the average difference in gene expression between normal and cancer for two classes of genes: hubs (defined as nodes of degree $>316$) and nodes of degree 1 (k=1). The number of each is indicated, and the P-value is from a Kolmogorov-Smirnov test, testing for a difference in their statistical distributions.
}
\label{fig:2}       % Give a unique label
\end{center}
\end{figure}

To gain insight as to why there is an increased signaling flux through hubs in cancer, we focused on the hub gene exhibiting the largest increase in the local entropy rate. This was the gene {\it BUB1} ({\bf Fig.2A}). A scatterplot of the expression values of {\it BUB1} and that of its neighbors (813 neighbors) in a representative normal sample versus the corresponding expression values in a representative cancer sample, demonstrates that most of the expression differences involve increases in gene expression, implicating both the hub itself as well as some of its neighbors ({\bf Fig.2B}). Thus, for the majority of neighbors of {\it BUB1}, the increased expression of {\it BUB1} will, according to the mass action principle, drive increased signaling through this hub. Indeed, for each one of {\it BUB1's} neighbors we ranked its neighbors according to the largest increase in gene expression, revealing that the original hub (i.e. {\it BUB1}) ranked among the top 2$\%$ centile for 99$\%$ of the hub neighbors ({\bf SI Appendix, fig.S1}). Interestingly, this effect was not unique to {\it BUB1} since high-degree hubs generally exhibited a significant skew towards increased gene expression in cancer ({\bf Fig.2C-D}).\\
Confirming the biological significance of these results, we reached very similar conclusions by repeating the above analysis in the independent Affymetrix gene expression data set of normal liver and liver cancer tissue \cite{Wurmbach2007} ({\bf SI Appendix, fig.S2-S3}). Thus, the increased entropy rate in liver cancer is driven mainly by the increased expression of the highest degree hubs in the PPI network.

\subsection*{Effect of cancer perturbations on signaling entropy}

That the highest degree genes show preferential expression increases in cancer ({\bf Fig.2C-D, SI Appendix fig.S3}) suggests an intricate link between network topology and differential expression. Confirming this further, in both liver expression sets we also observed that the genes exhibiting the largest, or most significant, decreases in expression preferentially mapped to low-degree nodes ({\bf Fig.2C-D, SI Appendix fig.S3-S4}).

\begin{figure}[ht]
\begin{center}
% For example, with the graphicx style use
\includegraphics[scale=0.6]{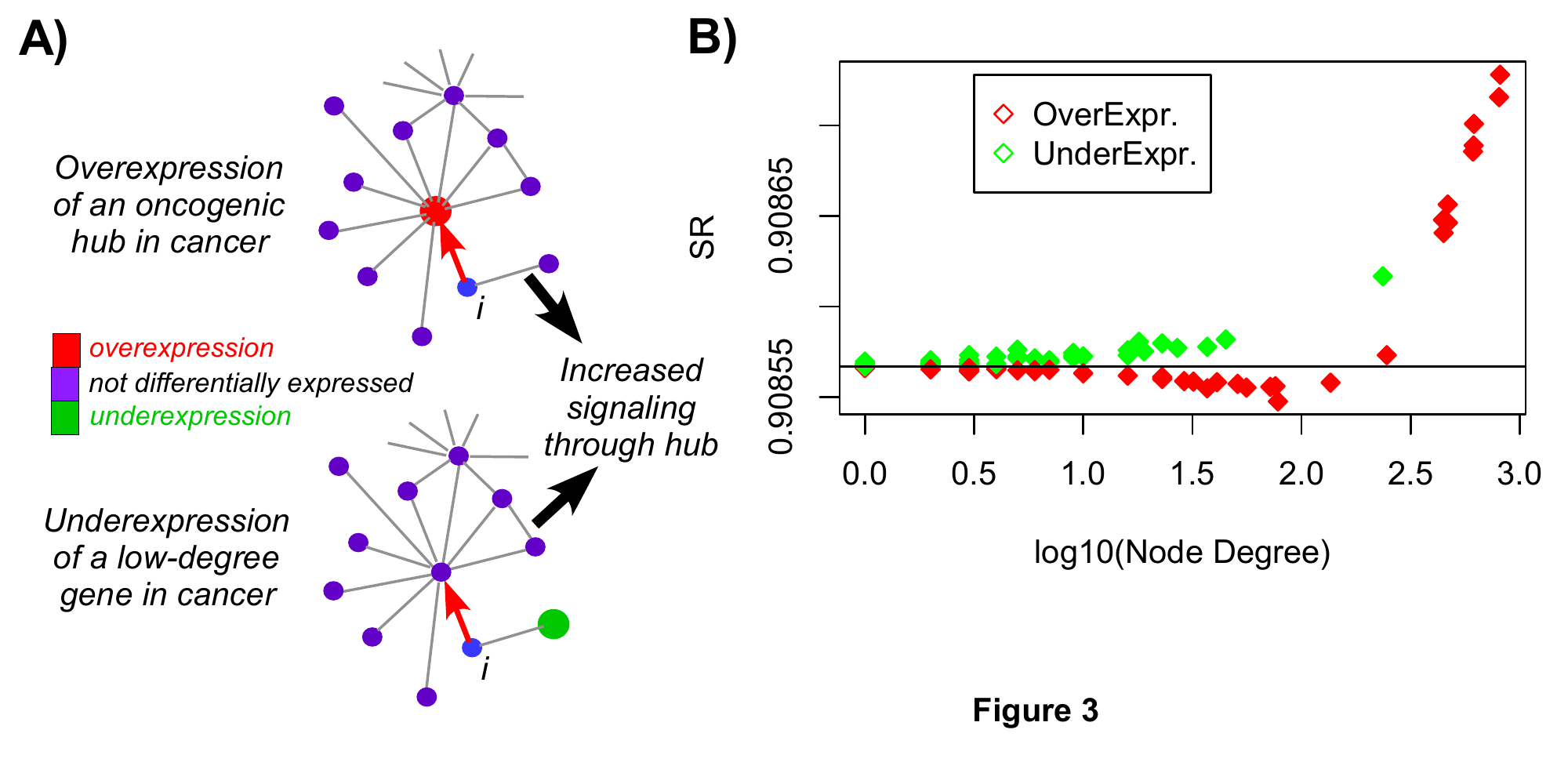}
%
% If no graphics program available, insert a blank space i.e. use
%\picplace{5cm}{2cm} % Give the correct figure height and width in cm
%
\caption{{\bf Effect of cancer perturbations on signaling entropy:} {\bf A)} Examples of two expression perturbations typically found in cancer. Top depicts the example of an oncogenic hub undergoing overexpression in cancer, which has the effect of drawing in signaling flux from a neighbour $i$. Example at the bottom depicts the underexpression of a low-degree ``tumour suppressor'' node (e.g. a transcripton factor), which from the perspective of node $i$ causes, indirectly, an increased signaling flux through the nearby hub. {\bf B)} Perturbation analysis of the top 100 genes ranked according to fold-change between normal and liver cancer. Plots shows the entropy rate after perturbation (y-axis) against node-degree (x-axis), with colors indicating over or underexpression. Black horizontal line defines the entropy rate of the average expression profile of normal liver (i.e. before the perturbation).
}
\label{fig:3}       % Give a unique label
\end{center}
\end{figure}

This intricate correlation between differential expression and node degree motivated us to pursue a deeper understanding of the complex interplay between network topology, gene expression perturbations and entropy rate. Intuitively, and from the perspective of a gene $i$ that interacts with an oncogenic hub, overexpression of the latter would lead to an increased outgoing signaling flux of node $i$ towards the hub, potentially leading to an increase in the overall entropy rate ({\bf Fig.3A}). Interestingly, underexpression of a low-degree node, which may connect to a hub either directly or indirectly through an intermediate node $i$ would also lead to an increased signaling flux through the hub ({\bf Fig.3A}). Thus, the two characteristic topological features of differential gene expression changes in cancer could synergize causing increased signaling flux through key hubs. To test whether this is indeed the case, we performed a perturbation analysis for the top 100 genes ranked according to fold-change between normal liver and liver cancer. The initial signaling distribution was defined by invoking the mass action principle on the average expression profile over all 50 normal liver samples. Next, each of the top 100 ranked genes was individually perturbed by changing its expression level according to the observed difference between normal and cancer tissue. Confirming our hypothesis, underexpressed genes (which generally did not target hubs) led to marginal increases in the entropy rate, whilst overexpressed hubs caused significant entropy increases ({\bf Fig.3B}). Interestingly however, overexpression led to marginal entropy decreases whenever it did not target the highest degree hubs, suggesting that such perturbations draw away signaling flux from the major hubs ({\bf Fig.3B}).

\subsection*{The effect of perturbations on signaling entropy is dependent on network topology}

To further investigate the effect of individual perturbations on signaling entropy, as well as the role of the underlying network topology, we devised a simulation framework on toy networks, perturbing each node in turn, and recording the effect on the entropy rate ({\bf Fig.4A, Appendix}). To simplify the analysis we considered an initial uniform edge weight configuration, defining an unbiased random walk on the graph. We note that this initial configuration represents a state of relatively high signaling entropy, {\it but not of maximal entropy} (see {\bf Appendix}). As activating perturbations we consider local increases in gene expression, whereby all the weights of edges converging on a perturbed node $i$ are assigned a relatively large weight ({\bf Fig.4B}). Thus, as seen from the perspective of a neighboring node $j$, before perturbation, node $j$ has maximal local entropy, given by $\log{k_j}$ (where $k_j$ is the degree of node $j$), whilst after the perturbation, the node's local entropy is close to 0 ({\bf Fig.4B}). We emphasize again that although in the initial configuration all local entropies are maximal, that the initial entropy rate over the whole network is not maximal (see {\bf Appendix}). Thus, after the perturbation, the global entropy rate of the network could increase or decrease.

\begin{figure}[ht]
\begin{center}
% For example, with the graphicx style use
\includegraphics[scale=0.6]{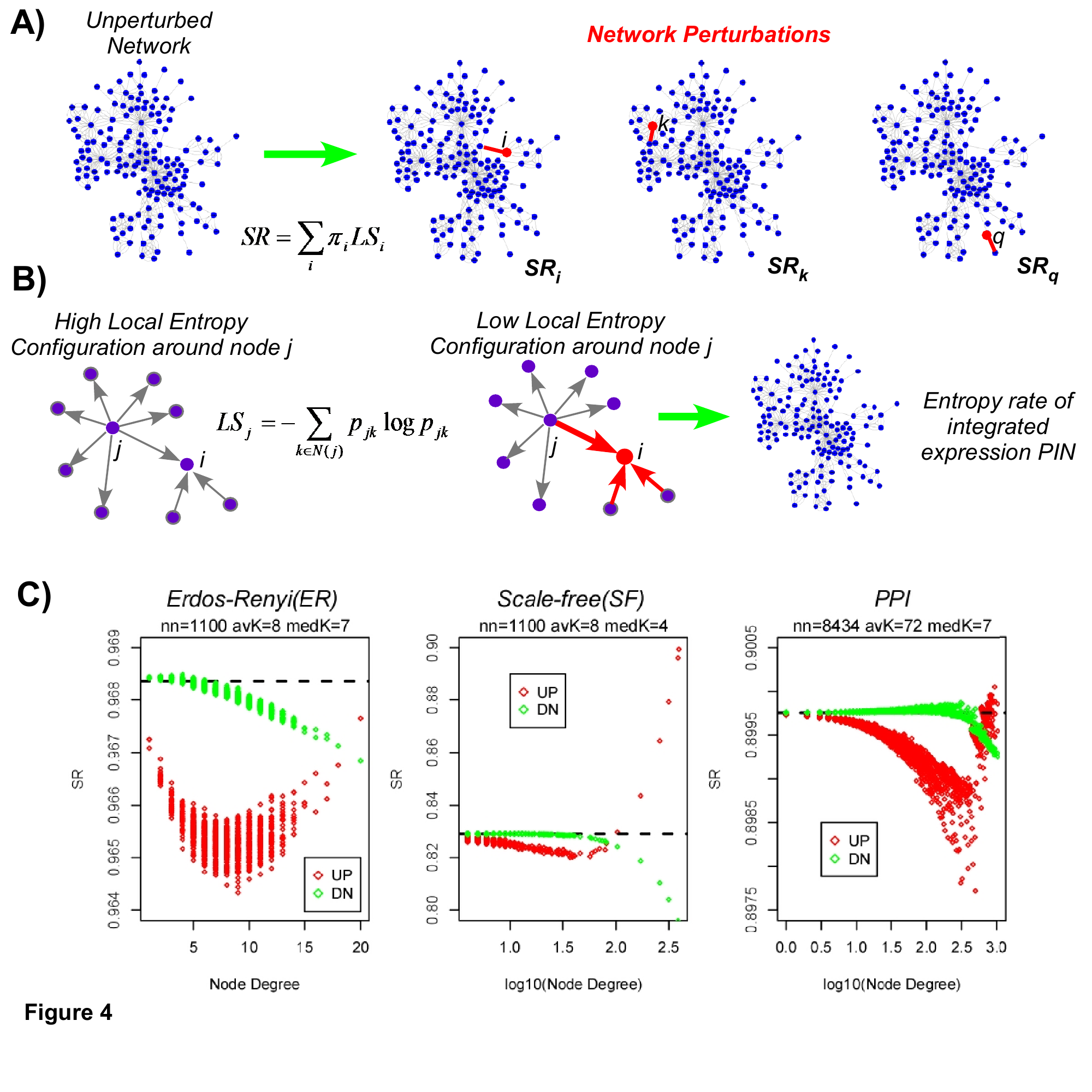}
%
% If no graphics program available, insert a blank space i.e. use
%\picplace{5cm}{2cm} % Give the correct figure height and width in cm
%
\caption{{\bf Cancer perturbations may increase the entropy rate on networks with scale-free topology but not on random Poisson graphs:} {\bf A)} A cartoon of the network perturbation analysis: each node $i$ of the network is perturbed in turn by changing its expression value. The case of overexpression is here indicated in red. The increased expression draws in signaling flux from neighbours (only one perturbed edge is shown). The entropy rate of the network after perturbing node $i$, $SR_i$, is computed and compared to the entropy rate $SR$ of the original unperturbed network. For $n$ nodes in the network we get a distribution of entropy rate changes $(SR_i-SR,i=1,...,n)$. {\bf B)} Zoomed-in version of a network perturbation, whereby a node $i$ undergoes a perturbation (here overexpression). From the perspective of a neighbouring node $j$, the perturbation causes a low signaling entropy configuration around node $j$. Key question is how does this perturbation affect the global entropy rate. {\bf C)} Perturbation analysis result, in which each node (gene) of the network was perturbed through overexpression (red) or underexpression (green). Plotted is the global entropy rate (SR) after the perturbation (y-axis) against the degree of the perturbed node (x-axis), for 3 different networks: Erdos-Renyi (ER) graph, scale-free (SF) network and the full PPI network (PPI). Black dashed line denotes the entropy rate before the perturbation. In each plot there as many data points as there are nodes in the network, each value corresponding to the perturbation of only one node. Number of nodes (nn), average degree (avK) and median degree (medK) are given.
}
\label{fig:4}       % Give a unique label
\end{center}
\end{figure}

In order to understand the potential impact of network topology, we first conducted the perturbation analysis above on Erdos-Renyi (ER) random graphs, for which the degree distribution is Poisson. For such ER graphs, we observed that activating perturbations (i.e. increases in gene expression), always led to a reduction in the global entropy rate, irrespective of node degree ({\bf Fig.4C}). Repeating the analysis for inactivating perturbations, i.e causing nodes to undergo underexpression, we observed that almost all nodes led to a decrease in entropy. Thus, given that cancer is characterised by an increase in signaling entropy, this suggests that the emergence of an increased signaling promiscuity regime in cancer must be due either to specific topological features not present in random graphs, or to non-random combinations of perturbations.\\
To investigate this further, we next performed the same perturbation analysis above, but now on networks characterised by a scale-free (or near scale-free) topology, a key feature of real biological networks \cite{Barabasi2004}. The scale-free networks were matched to the same size and average connectivity than the previously considered Erdos-Renyi graphs. Remarkably, in scale-free networks we observed that activating perturbations exhibited a bi-modal response, with perturbations at lower-degree nodes resulting in a reduction of the global entropy rate, whilst hubs exhibited increases ({\bf Fig.4C}). In fact, we observed two distinct regimes with an opposite functional relationship between entropy change and node-degree ({\bf Fig.4C}). In the low-degree regime, the entropy rate decreased as node degree increases, whereas in the high-degree regime one observes entropy increases ({\bf Fig.4C}). Interestingly, this bi-phasic behaviour was not seen for inactivating perturbations where we observed a monotonic decrease of entropy with node degree ({\bf Fig.4C}). In stark contrast to Poisson networks, high-degree nodes in the scale-free network exhibited a bi-modal response dependent on the directionality of the perturbation ({\bf Fig.4C}): overexpressed hubs led to entropy increases, while underexpressed hubs led to corresponding decreases.\\
Next, we wanted to test whether this bi-phasic and bi-modal behaviour is also seen in real PPI networks. We first checked that our PPI network exhibited an approximate scale-free topology ({\bf SI Appendix, fig.S5}). Its clustering coefficient was also significantly higher than that of a degree-distribution matched scale-free network ({\bf SI Appendix, fig.S5}). Performing the perturbation analysis on the PPI network, we observed once again two phases, which was particularly striking for activating perturbations, with one phase exhibiting a negative correlation between node degree and entropy, whilst the hub regime exhibited a positive correlation ({\bf Fig.4C}). Very interestingly, however, increases in entropy were only observed for the highest-degree hubs, with lower-degree hubs exhibiting decreases which were surprisingly also of a larger magnitude ({\bf Fig.4C}). Thus, in networks with a scale-free or an approximate scale-free topology, overexpression of the highest degree hubs leads to an increase in the entropy rate. But increasing signaling flux through lower-degree nodes, even if of relatively high degree, leads to an overall reduction in the diffusion rate.\\
From the combined perturbation analysis, we can thus see that individual perturbations on an Erdos-Renyi graph, be they activations or inactivations (but both causing a local reduction in entropy), invariably lead to  a reduction in the global entropy rate. This is in stark contrast to networks with a scale-free or approximate scale-free topology, where we observe that gene activations can have opposite effects on entropy rate depending on the degree of the activating nodes.

\begin{figure}[ht]
\begin{center}
% For example, with the graphicx style use
\includegraphics[scale=0.6]{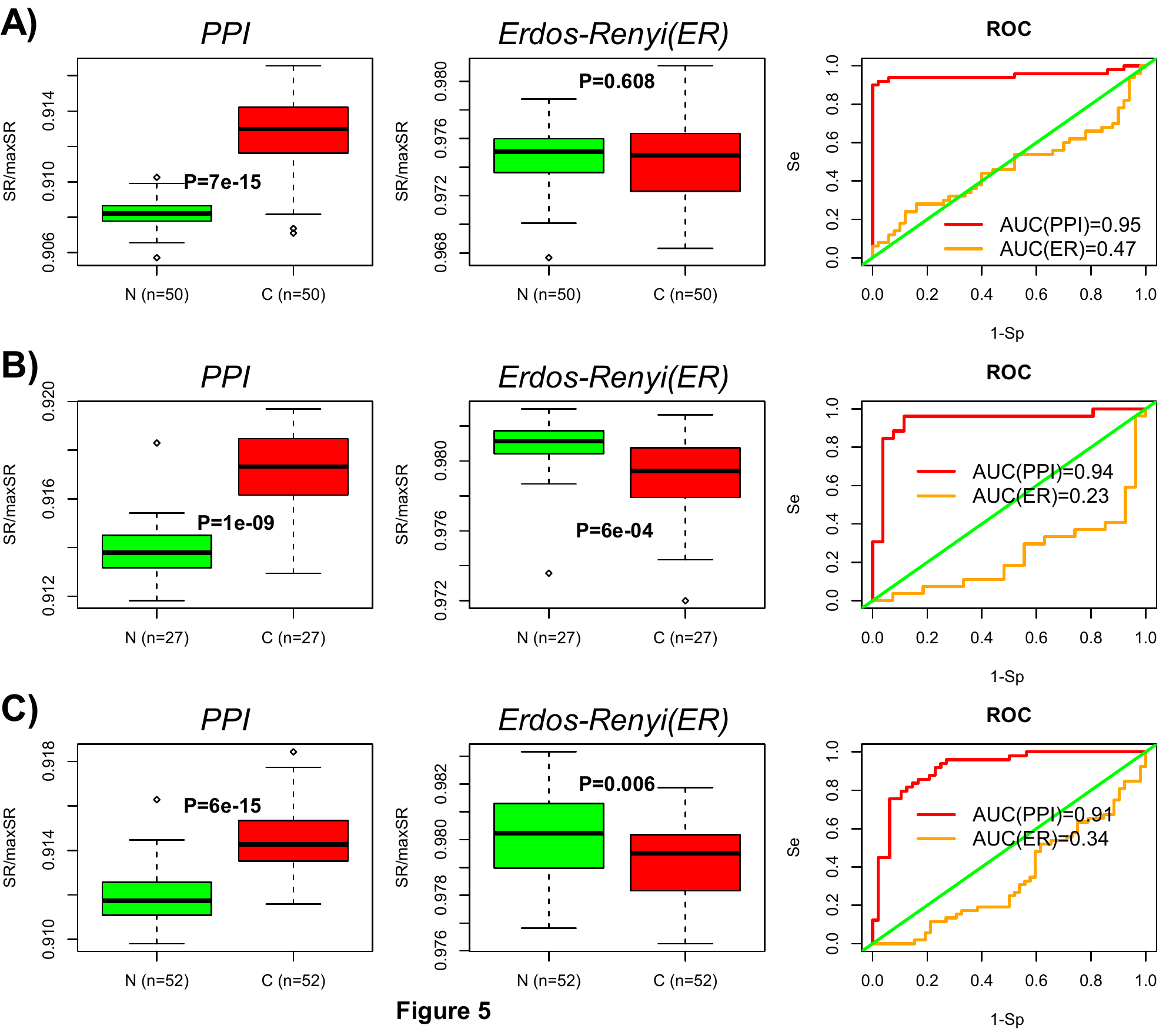}
%
% If no graphics program available, insert a blank space i.e. use
%\picplace{5cm}{2cm} % Give the correct figure height and width in cm
%
\caption{{\bf Entropy rate increase in cancer requires the scale-free topology of the PPI network:} {\bf A)} Boxplots of the entropy rate (SR) for the 50 normal liver and 50 liver cancer samples as evaluated on the original full PPI network (left), as well as on an equivalent Erdos-Renyi graph (middle). P-values are from a Wilcoxon-rank sum test. Corresponding ROC curves and AUC values (right). {\bf B)} As A) but for TCGA RNA-Seq data from 27 colon cancers and 27 matched normals. {\bf C)} As A) but for TCGA RNA-Seq data from 52 prostate cancers and 52 matched normals.
}
\label{fig:5}       % Give a unique label
\end{center}
\end{figure}

\subsection*{Entropy rate increase in cancer requires a scale-free interaction network topology}
The previous perturbation analysis strongly supports the view that a scale-free, or near scale-free network topology, is important for the observed increased entropy rate in cancer. To test this formally, we recomputed the entropy rate of all 50 normal liver and 50 liver cancer samples, but now using an underlying Erdos-Renyi (ER) interaction network matched to the same size and average connectivity of the full PPI network. In order to faithfully preserve the correlation between gene expression and node degree of the PPI network, nodes of the ER network were ranked according to degree and gene expression values assigned according to their corresponding rank/centile in the original PPI network. Thus, this node mapping between the two networks preserves the observed rank correlation between differential expression and node-degree, allowing us to objectively assess the importance of the scale-free property. Recomputation of the entropy rates of all 100 samples on the ER network revealed no significant difference between normal and cancer, thus demonstrating that the observed entropy rate increase in cancer requires the scale-free property of the interaction network ({\bf Fig.5A}). Supporting this further, we observed, in two other matched normal-cancer RNA-seq expression sets from the TCGA, that the entropy was no longer higher in cancer when the PPI network was replaced with an equivalent ER graph ({\bf Fig.5B-C}). In independent Affymetrix gene expression data, we observed that the cancer-associated increase in the entropy rate was reduced upon computing entropy on an equivalent ER network, in three out of four studies ({\bf SI Appendix, fig.S6}). Thus, in 6/7 data sets, there was a reduction in the entropy rate difference between cancer and normal tissue (Binomial, P=0.008), supporting the view that a scale-free interaction topology is indeed necessary for the higher entropy signaling dynamics of cancer.

\section*{Discussion}

Signaling entropy, a measure of the overall uncertainty or promiscuity in signaling patterns within a cellular sample, has been shown to be of biological significance in a variety of different contexts \cite{West2012,Banerji2013,Teschendorff2014}. In cellular differentiation it provides a proxy to the energy potential (i.e. height) of Waddington's epigenetic landscape, allowing the differentiation potential of a sample to be assessed purely from its genome-wide transcriptomic profile \cite{Banerji2013}. Similarly, signaling entropy also provided us with a useful framework in which to identify specific systems-level features characterising cancer, one of which being the increased signaling promiscuity of cancer compared to its corresponding normal tissue \cite{West2012,Banerji2013,Teschendorff2014}. This is important because an increased signaling promiscuity could underlie the increased phenotypic plasticity of cancer, as observed e.g. by Pisco et al \cite{Pisco2013}.\\
In this work we aimed to obtain a deeper theoretical understanding as to (i) why signaling entropy is increased in cancer and (ii) why it is such a robust discriminatory feature. We have here demonstrated that the increase in signaling entropy is driven by two factors. First, a subtle positive correlation between differential gene expression and the degree of the corresponding proteins in the PPI network. This correlation amounts to hubs exhibiting preferential increases in gene expression, whilst those genes exhibiting the most significant underexpression map preferentially to low-degree nodes. Second, the observed increase of entropy in cancer requires the scale-free (or near scale-free) topology characterising PPI networks. Indeed, by considering a Poisson network with an identical rank correlation coefficient between differential expression and node-degree, we no longer consistently observed a significant increased entropy rate in cancer ({\bf Fig.5}). Given the demonstrated biological significance of the entropy rate \cite{Banerji2013,Teschendorff2014}, this last result thus exposes a deep connection between the cancer phenotype and the underlying scale-free property of real PPI networks. It suggests that if the degree distribution of a PPI network were Poisson, that the transcriptomic changes seen in cancer would not define a highly promiscuous signaling regime. In other words, our data support the view that cancer ``hijacks'' the scale-free property of real signaling networks in order to facilitate increased signaling promiscuity and intra-tumour heterogeneity.\\
The novel insights described above also explain why the entropy rate provides such a robust discriminatory feature of the cancer phenotype. The robustness stems from the subtle correlation between differential expression and node-degree. Although gene expression data is notoriously noisy, there is generally speaking good agreement across independent studies when comparing the changes in differential gene expression between two marked phenotypes such as normal and cancer tissue \cite{Rhodes2004}. Secondly, although current PPI networks only represent mere caricatures of the real interactions in a cell, the ``hubness'' of a protein is likely to be a very robust feature. Indeed, that a given protein has exceptionally many interactions, thus defining a hub in a network, is likely to be a very robust feature, despite the fact that the specific interaction space of the hub may contain many false negatives and false positives \cite{Cerami2011}. Thus, the relative robustness of differential expression and hubness drives the robustness of the observed correlation between differential expression and node degree, which in turn explains why increased signaling entropy is such a consistent feature of the cancer phenotype \cite{West2012,Banerji2013}. Given the robustness of signaling entropy as a marker of differentiation potency \cite{Banerji2013}, it is therefore tempting to speculate that a subtle correlation between differential expression and node degree also exists in the context of normal cellular differentiation. Furthermore, it will be interesting to explore if the scale-free or near scale-free topology of PPI networks is also a key element underlying the nature of pluripotency, multipotency and terminal differentiation.\\
Although many previous studies have explored differential gene expression changes in cancer and other diseases in relation to network topology \cite{Lu2007,Platzer2007,Komurov2010bmc,Komurov2010,Jonsson2006,Tuck2006,Hudson2009,Ulitsky2007,Chuang2007,Yu2007}, most of these have either focused on global topological properties, or on finding differential gene modules, or on studying {\it absolute} changes in differential expression. Indeed, a number of studies agree in reporting that absolute differential expression correlates negatively with node degree, meaning that hubs exhibit, on the whole, much smaller changes in expression between disease phenotypes \cite{Lu2007,Komurov2010bmc}. Interestingly, however, relatively little attention has been paid to studying the {\it directionality} of differential gene expression in cancer in relation to node degree. Here we have shown that there exists a subtle yet significantly positive correlation between differential expression and protein-degree. On its own, the biological significance of this correlation is unclear. However, by interpreting this correlation in the novel contextual framework of signalling entropy, we have here shown how, in the context of real (near) scale-free networks, it could underpin the increased phenotypic plasticity of cancer.\\
In summary, increased expression of oncogenic hubs, as well as reduced expression of network-peripheral tumour suppressor genes, in interaction networks characterised by a (near) scale-free topology, drives the high signaling entropy of cancer and could thus underpin cancer's phenotypic robustness and plasticity. Further in-depth study of the complex interplay between local protein activity changes, their interaction network topology and the effect on signaling entropy is warranted.

\section*{Appendix}

\subsection*{The protein protein interaction (PPI) network}
We used a PPI network similar to that used in our previous publication \cite{West2013}. Briefly, the human interaction network derives from the Pathway Commons Resource ({\it www.pathwaycommons.org})\cite{Cerami2011}, which brings together protein interactions from several distinct sources, including the Human Protein Reference Database (HPRD) \cite{Prasad2009}, the National Cancer Institute Nature Pathway Interaction Database (NCI-PID) ({\it pid.nci.nih.gov}), the Interactome (Intact) {\it http://www.ebi.ac.uk/intact/} and the Molecular Interaction Database (MINT) {\it http://mint.bio.uniroma2.it/mint/}. Protein interactions in this network include physical stable interactions such as those defining protein complexes, as well as transient interactions such as post-translational modifications and enzymatic reactions found in signal transduction pathways, including 20 highly curated immune and cancer signaling pathways from NetPath ({\it www.netpath.org}) \cite{Kandasamy2010}. The network focuses on non-redundant interactions, only included nodes with an Entrez gene ID annotation and on the maximally connected component thereof, resulting in a connected network of 8,434 nodes (unique Entrez IDs) and 303,600 documented interactions. 

\subsection*{Normal and cancer tissue gene expression data sets}
We focused on liver cancer because the associated normal tissue constitutes a relatively homogeneous mass of cells, and thus the entropy rate is less likely to be influenced by changes in tissue-type composition. We downloaded the level 3 gene normalized RNA-Seq data from the TCGA ({\it www.cancergenome.nih}) for a matched subset of 50 normal liver and 50 liver cancer samples. As validation, we considered an Affymetrix expression data set, consisting of 37 normal livers (including normal liver, cirrhosis and dysplasia) + 38 liver cancers \cite{Wurmbach2007}. To test generalisability, we also downloaded level 3 RNA-Seq gene normalised data from the TCGA for prostate cancer (52 cancers \& 52 matched normals) and colon cancer (27 cancers \& 27 matched normals). The other normal/cancer Affymetrix expression sets used have been described previously \cite{Banerji2013}.

\subsection*{Construction of the sample specific stochastic matrix and entropy rate}
The construction of the entropy rate follows the same method described in our earlier work \cite{Banerji2013,Teschendorff2014}. Briefly, we use the mass action principle to define a stochastic matrix, $p_{ij}$, for each individual sample. In detail, let $E_i$ denote the normalised expression level of gene $i$ in a given sample. For a given neighbour $j\in N(i)$ (where $N(i)$ labels the neighbours of $i$ in the PPI), the mass-action principle means that the probability of interaction with $j$ is approximated by the product $E_{i}E_{j}$, i.e. $p_{ij}\propto E_{i}E_{j}$. Normalising this to ensure that $\sum_{j}p_{ij}=1$, we get for the stochastic matrix,
\begin{equation}
p_{ij}=\frac{E_{j}}{\sum_{k\in N(i)}E_{k}} \qquad\forall j\in N(i)
\label{eq:stochm}
\end{equation}
Clearly, if $j\notin N(i)$, then $p_{ij}=0$. From this stochastic matrix one can then construct a local signaling entropy ($LS$) as
\begin{equation}
LS_i = -\sum_{j\in N(i)}{p_{ij}\log{p_{ij}}}
\end{equation}
which reflects the level of uncertainty or redundancy in the local interaction probabilities. We note that the above expression for the local entropy is not normalised so that the maximum possible entropy depends on the degree ($k_i$) of the node. In fact,
\begin{equation}
\max{LS_i}=\log{k_i}
\end{equation}
Finally, the signaling entropy rate, $SR$, is defined in terms of the stationary distribution (or invariant measure) $\pi$ of the stochastic matrix ($\pi p = \pi$), as \cite{Latora1999,GomezGardenes2008}
\begin{equation}
SR=\sum_i{\pi_iLS_i}
\end{equation}
i.e. this global signaling entropy rate is a weighted average of the local entropies $LS_i$. We note that although $LS_i$ is independent of the expression level of gene $i$, that the gene's contribution to the entropy rate, i.e. $\pi_iLS_i$, is not. This is because $\pi_i$ will depend on the gene $i$'s expression level. In this work we refer to the term $LSR_i\equiv\pi_iLS_i$ as the local entropy rate of gene $i$, whereas $LS_i$ is just the gene $i$'s local entropy.

\subsection*{The maximum entropy rate}
Given a connected network, the maximum entropy rate, $maxSR$, over the network does not depend on the gene expression data but only on the adjacency matrix of the network. In fact, the maximum entropy rate is attained for a stochastic matrix $p_{ij}$ given by \cite{Demetrius2005}
\begin{equation}
p_{ij}=\frac{A_{ij}v_j}{\lambda v_i}
\end{equation}
where $v$ and $\lambda$ are the dominant right eigenvector and eigenvalue of the adjacency matrix $A$, respectively. Thus, it is important to note that the configuration of maximal local entropy, i.e. the configuration where for each node $i$, $p_{ij}=A_{ij}/k_i$ and $LS_i=\log{k_i}$, is not the configuration of maximal global entropy.

\subsection*{Perturbation simulation analysis}
In what follows we describe the perturbation analysis performed on Erd\"os-Renyi and scale-free networks, as well as on the full real PPI network described earlier. The calculation of the global signaling entropy rate is simplified significantly by the fact that the stochastic matrix defined by equation~\ref{eq:stochm} has the detailed balance property, i.e. the stationary distribution obeys not only $\pi p = \pi$, but the more restrictive condition $\pi_i p_{ij} = \pi_j p_{ji}$. This detailed balance condition can be shown to imply
\begin{equation}
\pi_i = \frac{1}{F} x_i x_{T,i}
\end{equation}
where F is a normalisation constant and $x_{T,i}=\sum_{j\in N(i)} x_j$.\\
The initial configuration for the perturbative analysis is that of maximal local entropy for each node in the network, which as explained previously, does not represent the state of global maximum entropy. To construct this initial configuration we set the expression level of each gene/node to be identical $x_i=x$. Thus, in the initial configuration, $x_{T,i}=k_ix$, and from detailed balance we obtain for the stationary distribution that
\begin{equation}
\pi_i=\frac{1}{F}x_ix_{T,i}=\frac{k_i}{V\bar{k}}
\end{equation}
where $V$ is the number of nodes in the network and where $\bar{k}$ is the average degree. As far as the entropy is concerned, the local entropy of each node $i$ is simply $\log{k_i}$, so the initial entropy rate is simply
\begin{equation}
SR_o=\frac{1}{V\bar{k}}\sum_{i}{k_i\log{k_i}}
\end{equation}
Now let us consider perturbing a gene in the network by altering its expression level by an amount $\lambda$. Without loss of generality we label the perturbed node by the index ``1'', so that after perturbation, the expression levels in the network are described by $x'_i=x+\delta_{i1}\lambda$. The new stationary distribution then becomes
\begin{eqnarray}
\pi'_1&\propto &(x+\lambda)k_1x \\
\pi'_i&\propto &x(x+\lambda+(k_i-1)x) \qquad \forall i\in N(1)\\
\pi'_i&\propto &k_ix^2 \qquad \forall i\notin N(1)\cup 1 
\end{eqnarray}
For the local entropies, we get
\begin{eqnarray}
LS'_i&=&LS_i \qquad \forall i\in N/N(1)\\
LS'_i&=& -\sum_{j\in N(i)/1}{p'_{ij}\log{p'_{ij}}}\\
    & & -p'_{i1}\log{p'_{i1}} \qquad \forall i\in N(1) 
\end{eqnarray}
where for $i\in N(1)$, $p'_{i1}=(x+\lambda)/(x+\lambda+(k_i-1)x)$ and $p'_{ij}=x/(x+\lambda+(k_i-1)x)$ ($j\neq 1$). Thus, the change in the entropy rate, $\Delta SR=SR'-SR_o$, is easily computable following any perturbation.\\
In the actual analysis, when performing activating perturbations, we set $x=2$ and $\lambda=14$, whilst, when modeling inactivating perturbations, we set $x=16$ and $\lambda=-14$. These values are typical for logged Affymetrix or Illumina data, with highly expressed genes normally exhibiting values larger than 12, and lowly expressed genes showing values smaller than 4.

\begin{acknowledgement}
AET is supported by the Chinese Academy of Sciences and the Max-Planck Society. 
\end{acknowledgement}
%
%\section*{Appendix}
%\addcontentsline{toc}{section}{Appendix}

%
\bibliographystyle{spmpsci}
%\bibliography{pertsrAET}

\begin{thebibliography}{10}
\providecommand{\url}[1]{{#1}}
\providecommand{\urlprefix}{URL }
\expandafter\ifx\csname urlstyle\endcsname\relax
  \providecommand{\doi}[1]{DOI~\discretionary{}{}{}#1}\else
  \providecommand{\doi}{DOI~\discretionary{}{}{}\begingroup
  \urlstyle{rm}\Url}\fi

\bibitem{Albert2000}
Albert, R., Jeong, H., Barabasi, A.L.: Error and attack tolerance of complex
  networks.
\newblock Nature \textbf{406}(6794), 378--82 (2000)

\bibitem{Ideker2010}
Bandyopadhyay, S., Mehta, M., Kuo, D., Sung, M.K., Chuang, R., Jaehnig, E.J.,
  Bodenmiller, B., Licon, K., Copeland, W., Shales, M., Fiedler, D., Dutkowski,
  J., Guénolé, A., {van Attikum}, H., Shokat, K.M., Kolodner, R.D., Huh,
  W.K., Aebersold, R., Keogh, M.C., Krogan, N.J., Ideker, T.: Rewiring of
  genetic networks in response to dna damage.
\newblock Science \textbf{330}(6009), 1385--1389 (2010)

\bibitem{Banerji2013}
Banerji, C.R., Miranda-Saavedra, D., Severini, S., Widschwendter, M., Enver,
  T., Zhou, J.X., Teschendorff, A.E.: Cellular network entropy as the energy
  potential in waddington's differentiation landscape.
\newblock Sci Rep \textbf{3}, 3039 (2013)

\bibitem{Barabasi1999}
Barabasi, A.L., Albert, R.: Emergence of scaling in random networks.
\newblock Science \textbf{286}, 509--512 (1999)

\bibitem{Barabasi2004}
Barabasi, A.L., Oltvai, Z.N.: Network biology: understanding the cell's
  functional organization.
\newblock Nat Rev Genet \textbf{5}(2), 101--113 (2004)

\bibitem{Califano2011}
Califano, A.: Rewiring makes the difference.
\newblock Mol Syst Biol \textbf{7}, 463 (2011)

\bibitem{Cerami2011}
Cerami, E.G., Gross, B.E., Demir, E., Rodchenkov, I., Babur, O., Anwar, N.,
  Schultz, N., Bader, G.D., Sander, C.: Pathway commons, a web resource for
  biological pathway data.
\newblock Nucleic Acids Res \textbf{39}(Database), D685--D690 (2011)

\bibitem{Chuang2007}
Chuang, H.Y., Lee, E., Liu, Y.T., Lee, D., Ideker, T.: Network-based
  classification of breast cancer metastasis.
\newblock Mol Syst Biol \textbf{3}, 140 (2007)

\bibitem{Creixell2012}
Creixell, P., Schoof, E.M., Erler, J.T., Linding, R.: {{N}avigating cancer
  network attractors for tumor-specific therapy}.
\newblock Nat. Biotechnol. \textbf{30}(9), 842--848 (2012)

\bibitem{Csermely2013}
Csermely, P., Korcsmaros, T.: Cancer-related networks: a help to understand,
  predict and change malignant transformation.
\newblock Semin Cancer Biol \textbf{23}(4), 209--12 (2013)

\bibitem{Demetrius2005}
Demetrius, L., Manke, T.: Robustness and network evolution-an entropic
  principle.
\newblock Physica A \textbf{346}, 682--696 (2005)

\bibitem{Dutkowski2011}
Dutkowski, J., Ideker, T.: Protein networks as logic functions in development
  and cancer.
\newblock PLoS Comput Biol \textbf{7}(9), e1002,180 (2011)

\bibitem{Erdos1959}
Erd\"os, P., Renyi, A.: On random graphs.
\newblock Pub Math \textbf{6}, 290--297 (1959)

\bibitem{GomezGardenes2008}
Gomez-Gardenes, J., Latora, V.: Entropy rate of diffusion processes on complex
  networks.
\newblock Phys Rev E Stat Nonlin Soft Matter Phys \textbf{78}(6), 065,102
  (2008)

\bibitem{Haynes2006}
Haynes, C., Oldfield, C.J., Ji, F., Klitgord, N., Cusick, M.E., Radivojac, P.,
  Uversky, V.N., Vidal, M., Iakoucheva, L.M.: Intrinsic disorder is a common
  feature of hub proteins from four eukaryotic interactomes.
\newblock PLoS Comput Biol \textbf{2}(8), e100 (2006)

\bibitem{Hudson2009}
Hudson, N.J., Reverter, A., Dalrymple, B.P.: A differential wiring analysis of
  expression data correctly identifies the gene containing the causal mutation.
\newblock PLoS Comput Biol \textbf{5}(5), e1000,382 (2009)

\bibitem{Ideker2012}
Ideker, T., Krogan, N.J.: Differential network biology.
\newblock Mol Syst Biol \textbf{8}, 565 (2012)

\bibitem{Jeong2001}
Jeong, H., Mason, S.P., Barabási, A.L., Oltvai, Z.N.: Lethality and centrality
  in protein networks.
\newblock Nature \textbf{411}(6833), 41--42 (2001)

\bibitem{Jonsson2006}
Jonsson, P.F., Bates, P.A.: Global topological features of cancer proteins in
  the human interactome.
\newblock Bioinformatics \textbf{22}(18), 2291--2297 (2006)

\bibitem{Kandasamy2010}
Kandasamy, K., Mohan, S.S., Raju, R., Keerthikumar, S., Kumar, G.S., Venugopal,
  A.K., Telikicherla, D., Navarro, J.D., Mathivanan, S., Pecquet, C.,
  Gollapudi, S.K., Tattikota, S.G., Mohan, S., Padhukasahasram, H.,
  Subbannayya, Y., Goel, R., Jacob, H.K., Zhong, J., Sekhar, R., Nanjappa, V.,
  Balakrishnan, L., Subbaiah, R., Ramachandra, Y.L., Rahiman, B.A., Prasad,
  T.S., Lin, J.X., Houtman, J.C., Desiderio, S., Renauld, J.C., Constantinescu,
  S.N., Ohara, O., Hirano, T., Kubo, M., Singh, S., Khatri, P., Draghici, S.,
  Bader, G.D., Sander, C., Leonard, W.J., Pandey, A.: Netpath: a public
  resource of curated signal transduction pathways.
\newblock Genome Biol \textbf{11}(1), R3 (2010)

\bibitem{Kim2014}
Kim, J., Vandamme, D., Kim, J.R., Munoz, A.G., Kolch, W., Cho, K.H.: Robustness
  and evolvability of the human signaling network.
\newblock PLoS Comput Biol \textbf{10}(7), e1003,763 (2014)

\bibitem{Komurov2010bmc}
Komurov, K., Ram, P.T.: Patterns of human gene expression variance show strong
  associations with signaling network hierarchy.
\newblock BMC Syst Biol \textbf{4}, 154 (2010)

\bibitem{Komurov2010}
Komurov, K., White, M.A., Ram, P.T.: Use of data-biased random walks on graphs
  for the retrieval of context-specific networks from genomic data.
\newblock PLoS Comput Biol \textbf{6}(8) (2010)

\bibitem{Latora1999}
Latora, V., Baranger, M.: Kolmogorov-sinai entropy rate versus physical
  entropy.
\newblock Phys Rev Lett \textbf{82}(3), 520--524 (1999)

\bibitem{Lu2007}
Lu, X., Jain, V.V., Finn, P.W., Perkins, D.L.: Hubs in biological interaction
  networks exhibit low changes in expression in experimental asthma.
\newblock Mol Syst Biol \textbf{3}, 98 (2007)

\bibitem{Pisco2013}
Pisco, A.O., Brock, A., Zhou, J., Moor, A., Mojtahedi, M., Jackson, D., Huang,
  S.: Non-darwinian dynamics in therapy-induced cancer drug resistance.
\newblock Nat Commun \textbf{4}, 2467 (2013)

\bibitem{Platzer2007}
Platzer, A., Perco, P., Lukas, A., Mayer, B.: Characterization of
  protein-interaction networks in tumors.
\newblock BMC Bioinformatics \textbf{8}, 224 (2007)

\bibitem{Prasad2009}
Prasad, T.S., Kandasamy, K., Pandey, A.: Human protein reference database and
  human proteinpedia as discovery tools for systems biology.
\newblock Methods Mol Biol \textbf{577}, 67--79 (2009)

\bibitem{Rhodes2004}
Rhodes, D.R., Yu, J., Shanker, K., Deshpande, N., Varambally, R., Ghosh, D.,
  Barrette, T., Pandey, A., Chinnaiyan, A.M.: Large-scale meta-analysis of
  cancer microarray data identifies common transcriptional profiles of
  neoplastic transformation and progression.
\newblock Proc Natl Acad Sci U S A \textbf{101}(25), 9309--9314 (2004)

\bibitem{Rolland2014}
Rolland, T., Tasan, M., Charloteaux, B., Pevzner, S.J., Zhong, Q., Sahni, N.,
  Yi, S., Lemmens, I., Fontanillo, C., Mosca, R., Kamburov, A., Ghiassian,
  S.D., Yang, X., Ghamsari, L., Balcha, D., Begg, B.E., Braun, P., Brehme, M.,
  Broly, M.P., Carvunis, A.R., Convery-Zupan, D., Corominas, R.,
  Coulombe-Huntington, J., Dann, E., Dreze, M., Dricot, A., Fan, C., Franzosa,
  E., Gebreab, F., Gutierrez, B.J., Hardy, M.F., Jin, M., Kang, S., Kiros, R.,
  Lin, G.N., Luck, K., MacWilliams, A., Menche, J., Murray, R.R., Palagi, A.,
  Poulin, M.M., Rambout, X., Rasla, J., Reichert, P., Romero, V., Ruyssinck,
  E., Sahalie, J.M., Scholz, A., Shah, A.A., Sharma, A., Shen, Y., Spirohn, K.,
  Tam, S., Tejeda, A.O., Trigg, S.A., Twizere, J.C., Vega, K., Walsh, J.,
  Cusick, M.E., Xia, Y., Barabasi, A.L., Iakoucheva, L.M., Aloy, P., a.~s. De,
  L., Tavernier, J., Calderwood, M.A., Hill, D.E., Hao, T., Roth, F.P., Vidal,
  M.: A proteome-scale map of the human interactome network.
\newblock Cell \textbf{159}(5), 1212--26 (2014)

\bibitem{Schramm2010}
Schramm, G., Nandakumar, K., Konig, R.: Regulation patterns in signaling
  networks of cancer.
\newblock BMC Syst Biol \textbf{4}(1), 162 (2010)

\bibitem{Serra-Musach2012}
Serra-Musach, J., Aguilar, H., Iorio, F., Comellas, F., Berenguer, A., Brunet,
  J., Saez-Rodriguez, J., Pujana, M.A.: Cancer develops, progresses and
  responds to therapies through restricted perturbation of the protein-protein
  interaction network.
\newblock Integr Biol (Camb) \textbf{4}(9), 1038--48 (2012)

\bibitem{Teschendorff2010bmc}
Teschendorff, A.E., Severini, S.: Increased entropy of signal transduction in
  the cancer metastasis phenotype.
\newblock BMC Syst Biol \textbf{4}, 104 (2010)

\bibitem{Teschendorff2014}
Teschendorff, A.E., Sollich, P., Kuehn, R.: Signalling entropy: A novel
  network-theoretical framework for systems analysis.
\newblock Methods  (2014)

\bibitem{Tuck2006}
Tuck, D.P., Kluger, H.M., Kluger, Y.: Characterizing disease states from
  topological properties of transcriptional regulatory networks.
\newblock BMC Bioinformatics \textbf{7}, 236 (2006)

\bibitem{Ulitsky2007}
Ulitsky, I., Shamir, R.: Identification of functional modules using network
  topology and high-throughput data.
\newblock BMC Syst Biol \textbf{1}, 8 (2007)

\bibitem{Wang2014}
Wang, S.J., Wang, Z., Jin, T., Boccaletti, S.: Emergence of disassortative
  mixing from pruning nodes in growing scale-free networks.
\newblock Sci Rep \textbf{4}, 7536 (2014)

\bibitem{West2013}
West, J., Beck, S., Wang, X., Teschendorff, A.E.: An integrative network
  algorithm identifies age-associated differential methylation interactome
  hotspots targeting stem-cell differentiation pathways.
\newblock Sci Rep \textbf{3}, 1630 (2013)

\bibitem{West2012}
West, J., Bianconi, G., Severini, S., Teschendorff, A.E.: Differential network
  entropy reveals cancer system hallmarks.
\newblock Sci Rep \textbf{2}, 802 (2012)

\bibitem{Wurmbach2007}
Wurmbach, E., Chen, Y.B., Khitrov, G., Zhang, W., Roayaie, S., Schwartz, M.,
  Fiel, I., Thung, S., Mazzaferro, V., Bruix, J., Bottinger, E., Friedman, S.,
  Waxman, S., Llovet, J.M.: Genome-wide molecular profiles of hcv-induced
  dysplasia and hepatocellular carcinoma.
\newblock Hepatology \textbf{45}(4), 938--947 (2007)

\bibitem{Yu2007}
Yu, H., Kim, P.M., Sprecher, E., Trifonov, V., Gerstein, M.: The importance of
  bottlenecks in protein networks: correlation with gene essentiality and
  expression dynamics.
\newblock PLoS Comput Biol \textbf{3}(4), e59 (2007)

\end{thebibliography}


%% References
%% Usage of  \cite is as follows:
%%    \cite{key}         ==>>  [#]
%%    \cite[chap. 2]{key} ==>> [#, chap. 2]
%%




\section*{Figures}




% \end{thebibliography}

\end{document}